\documentclass[aps,onecolumn,eqsecnum]
{revtex4}

\usepackage{natbib}
\usepackage{dcolumn}
\usepackage{amsmath}
\usepackage{graphics}
\usepackage{color}

\begin{document}


\title{Random Lasing in an Inhomogeneous and Disordered \\ System of Cold Atoms}

\author{\firstname{L.~V.}~\surname{Gerasimov}}
\author{\firstname{D.~V.}~\surname{Kupriyanov}}
\affiliation{Department of Theoretical Physics, Peter the Great St.~Petersburg Polytechnic University, 195251 St-Petersburg, Russia}%

\author{\firstname{M.~D.}~\surname{Havey}}
\affiliation{Department of Physics, Old Dominion University, Norfolk, VA 23529, USA}%

\email[]{kupr@dk11578.spb.edu}


\begin{abstract}
\textbf{Abstract}: We consider light trapping in an amplifying medium consisting of cold alkali-metal atoms; the atomic gas plays a dual role as a scattering and as a gain medium. We perform Monte-Carlo simulations for the combined processes. In some configurations of the inhomogeneous distribution this leads to a point of instability behavior and a signature of random lasing in a cold atomic gas.
\end{abstract}

\maketitle

\section*{Introduction}

Radiation trapping \cite{Molish}, light localization \cite{Anderson} and cooperative phenomena \cite{Dicke} involving coherent matter-radiation couplings are the subjects of various and intensive discussions in the scientific literature.  Among areas of current examination are exploration of super- and sub-radiant configurations \cite{Guerin,Yosuke,Tanya,Scully1}, cooperative effects including the cooperative Lamb shift \cite{Friedberg1,Manassah1,Rohlsberger1,Keaveney}, and the role of collective phenomena in dense and cold atomic gases \cite{Bienaime,Balik,Kemp,Pellegrino,Javanainen}.

In this paper, we concerned mainly the intriguing phenomenon of the so-called random laser \cite{Letokhov68,HuiCao,Hui2,Hui3} predicted originally by V. Letokhov in 1968. As was shown in his seminal paper \cite{Letokhov68}, if the amplifying active medium is distributed throughout another elastically scattering medium, and if the sample size $r_0$ fulfills the threshold condition $r_0>\pi\left(l_{\mathrm{tr}}l_g/3\right)^{1/2}$, where $l_{\mathrm{tr}}$ is the transport length of light diffusion and $l_g$ is the gain length, there is an instability point in the amplification process. Of course, in a realistic scenario the infinite amplification is impossible. If the sample volume is large enough such that this threshold is overcome then the amplification initiates a saturating feedback of radiation with the active medium and finally the radiation of the entire system can approach a steady state regime. The role of randomly distributed elastic scatterers in this process consists of effective trapping of the amplified radiation such that the scatterers serve as the cavity in a conventional laser scheme.

There have been many studies of random lasing, and signatures of the threshold behavior of the random lasing phenomenon in condensed matter systems. For instance, a narrow spectral feature was observed in the emission of metal ions distributed in powder paint \cite{Markushev}. Another example is the radiation of a dye active medium, which can be diffusely trapped by TiO${}_2$ microparticles distributed in the dye solution \cite{Lawandy}. These and other examples of laboratory manifestations of the phenomenon are reviewed in \cite{HuiCao}. The physical evidence of lasing from the disordered system is probably also observable in astrophysics as stimulated emission in certain stellar objects \cite{LetokhovJohansson}. In laboratory realizations it seems of special interest to systems with a quasi one dimensional geometry where the lasing in a disordered medium can be accompanied by localization phenomenon \cite{MilnerGenack,Matos}.

However, in the gas phase (e.g. atomic or molecular gases) unambiguous demonstration of all expected features of random lasing is still a challenging problem for experimental verification. It has been proposed that this effect could be observed in a system of cold alkali-metal atoms prepared in a magneto optical trap \cite{KaiserRandLaser1}. The unique statement formulated in that paper is that the radiation can be emitted and trapped by the same medium, which is active and scattering simultaneously.  Such a system can consist of billions atoms at the temperature $\sim$ $\mu K$ with very large optical thickness. The authors made their estimates based on a two-level model of optical transition pumped by a strong coherent field, i.e., for the Mollow-type system and they predicted an optical depth around several hundreds to approach the threshold condition. But as was recognized later, the complicated multilevel structure of an alkali-metal atom would be even more helpful in realization of the random lasing conditions. Indeed in the alkali-metal atom system the atoms can be repopulated among different hyperfine sublevels and then can be considered as distinguishable fractions of the matter. The progress in architectures of cold atomic systems is rapidly developing and the required conditions under a variety of circumstances may be attainable in the laboratory in the near future. Recently, elegant experimental results showing that the threshold conditions can be overcome in the system of cold atoms have been reported in \cite{KaiserRandLaser2,KaiserRandLaser3}.

In the present paper we discuss a particular and different scheme showing how both amplification and light trapping can be organized in such a system.  For this scheme, we show that threshold conditions can be attained, leading to a regime of random laser generation. We also briefly discuss how one component of such a scheme, an optically induced and quasi one dimensional soft cavity may be generated.

\section*{Amplification of the spontaneous Raman emission via\protect\\ radiation trapping mechanism}

\begin{figure}[tp]
{$\scalebox{0.8}{\includegraphics*{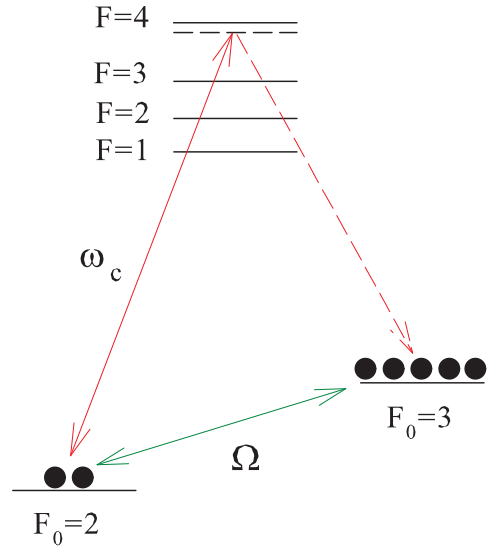}}$ }
\caption{(Color online) Energy levels and excitation diagram for the Raman process initiated in an ensemble of ${}^{85}$Rb via the $F_0= 2 \rightarrow F=3,2 \rightarrow F_0 = 3 $ transitions. The Raman emission is a result of the simultaneous action of microwave $\Omega$ and optical $\omega_c$ excitations, and both are linearly polarized along the quantization direction. The emitted light is partially trapped on the closed $F_0 = 3 \to F = 4$ transition in the optically thick atomic ensemble. While propagating through the atomic sample this light stimulates additional Raman emission and may increase the output fluorescence emerging from the sample.}
\label{fig1}%
\end{figure}

\begin{figure}[tp]
{$\scalebox{0.8}{\includegraphics*{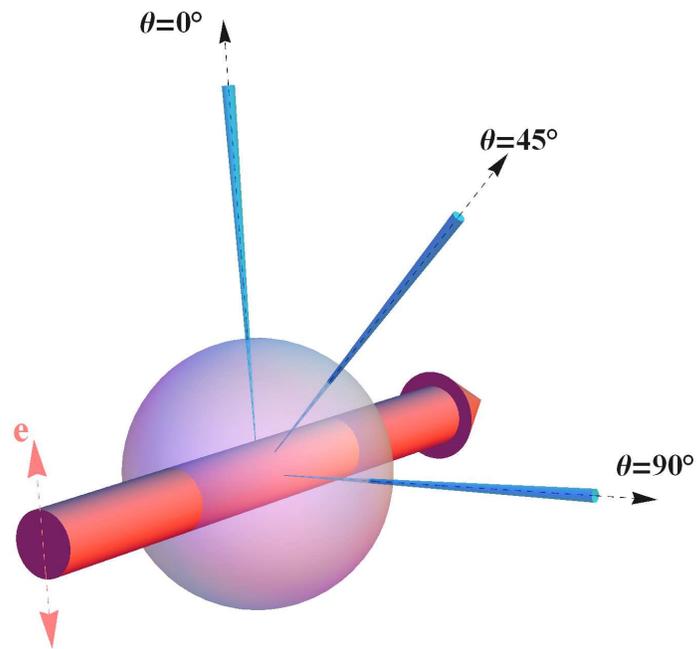}}$ }
\caption{(Color online) Excitation geometry: the linearly polarized control mode (shown as a pink arrow-tube) and microwave field (not shown) initiate the Raman emission for the process shown in Fig.~\ref{fig1}. The observation channels are parameterized by the polar angle $\theta=0^\circ,\,45^\circ,\,90^\circ$ from the control mode polarization direction.}
\label{fig2}%
\end{figure}

In this section we consider the mechanism of Raman-type amplification for radiation trapped in a disordered atomic medium. Such a mechanism can be organized in the $D_2$-line hyperfine manifold of any alkali-metal atom and in Fig.~\ref{fig1} we show ${}^{85}$Rb as a relevant example.   The respective excitation geometry is reproduced in Fig.~\ref{fig2}. Let the rubidium atoms initially populate the upper hyperfine sublevel in the ground state. Then some portion of them can be transferred to the lower sublevel with a microwave field. Upon applying a strong coherent field near resonance with the electric-dipole forbidden $F_0=2\to F=4$ transition one can initiate the off resonant Raman process via $F=3,2$ upper states, thus transferring atoms back to the $F_0=3$ sublevel. The radiation so generated will have a frequency nearly degenerate with the practically closed $F_0=3\to F=4$ transition.  The crucial peculiarity of such an excitation scheme is that the light emitted in the spontaneous Raman process can be strongly trapped by elastic scattering on the closed $F_0=3\to F=4$ transition if the optical depth of the sample is large. Then the portion of light, originally created in the spontaneous Raman process and then propagating via multiple scattering  through the disordered atomic ensemble can stimulate additional Raman emission. If we imbalanced the trapping conditions in such way that the amplification of the transported light overcome its losses, then we could arrive at the situation of a random lasing mechanism as discussed in \cite{GerasimovRandLaser}.  We briefly summarize that result below.

The above conditions justify that only a small portion of the atoms are repopulated from the upper hyperfine sublevel of the ground state and the stimulated amplification is expected to take place at a longer path length than the diffusion transport length. Then the amplification process can be subsequently considered in the perturbation theory and conveniently described in the diagram framework beyond any simplifications associated with light transport or diffusion equation. The electric field correlation function can be given by the sum
\begin{eqnarray}
D^{(1)}_{11'}(\omega)&=&\int_{-\infty}^{\infty} \exp(+i\omega\tau)\left\langle \hat{E}_{\mu'_1}^{(-)}(\mathbf{r}'_1,t-\tau)\hat{E}_{\mu_1}^{(+)}(\mathbf{r}_1,t)\right\rangle dt ,
\nonumber\\ \nonumber\\
D^{(1)}_{11'}(\omega)&=&D^{(\mathrm{0})}_{11'}(\omega)+D^{(\mathrm{1})}_{11'}(\omega)+D^{(\mathrm{2})}_{11'}(\omega)+\ldots
\label{Eq1}
\end{eqnarray}
where $\hat{E}_{\mu}^{(\pm)}(\mathbf{r},t)$ are the Heisenberg operators of the positive/negative frequency components of the electric field. For the sake of brevity we incorporated all the notation details into the numerical arguments such that $1\equiv \mu_{1},\mathbf{r}_1, \ldots$ etc. and indicate by the numbers the different arguments.
At each $n$-th scattering step the dynamics of the field obeys the following Bethe-Salpeter-type transformation for the correlation function of light accumulated in the previous $n-1$ scattering sequence:
\begin{eqnarray}
\lefteqn{D^{(n)}_{11'}(\omega)\propto \sum_{\begin{array}{c} \scriptstyle m_2=m'_2 \\ \scriptstyle \in F_0=2,3 \end{array}}%
\sum_{\begin{array}{c} \scriptstyle m_3,m'_3 \\ \scriptstyle \in F_0=3 \end{array}} \bar{\rho}_{m_3,m'_3}\times} %
\nonumber\\%
&&{\cal D}^{(R)}_{12}(\omega)\!\ast\!\alpha_{23}(\omega)\!\ast\! D^{(n-1)}_{33'}(\omega)\!\ast\! \alpha^{\dagger}_{3'2'}(\omega)\!\ast\! {\cal D}^{(A)}_{2'1'}(\omega),%
\label{Eq2}
\end{eqnarray}
where asterisks denote here the integral over spatial (if necessary) and sum over polarization and internal atomic variables. The zero order is associated with the light originally emitted in the spontaneous Raman process. Here ${\cal D}^{(R)}_{12}(\omega)$ and ${\cal D}^{(A)}_{12}(\omega)$ are the retarded and the advanced-type photon propagators, which express each other by Hermitian conjugation such that ${\cal D}^{(A)}_{2'1}={\cal D}^{(R)\ast}_{12'}$. The internal part of the transformation is the matrix product of the correlation function for light incident on a random atom with the scattering tensor $\alpha_{23}(\omega)$ and its Hermitian counterpart $\alpha^{\dagger}_{3'2'}(\omega)$ weighted with the density matrix $\bar{\rho}_{m_3,m'_3}$ of atoms populating the state $F_0=3$ (Fig.~\ref{fig1}). The sign of proportionality in the transformation (\ref{Eq2}) indicates any required factors associated with relevant scaling and geometry.

The light propagation dynamics mostly depends on the following kinetic parameters responsible for the light transport. The imaginary part of the sample susceptibility determines the extinction length for the plane wave entering the sample at the $e^{-1}$ level of losses. The inverse extinction length is given by
\begin{equation}
l_{\mathrm{ex}}^{-1}(\omega)=n_0\sigma_{\mathrm{ex}}(\omega)=4\pi k\, \mathrm{Im}\chi(\omega)\, ,%
\label{Eq3}
\end{equation}
where $n_0$ is a typical local atomic density and for the sake of simplicity we omitted in this estimate its spatial dependence. In an anisotropic sample this quantity can be defined only for a plane wave propagating along a specific direction associated with the main reference frame, such that $\chi(\omega)$ in Eq. (\ref{Eq3}) can be any of the major components of the susceptibility tensor. Otherwise the definition has to depend on the propagation direction and track the polarization properties of light. It is important to recognize that in the case of amplification the extinction length accumulates not only losses but also the gain associated with the stimulated Raman scattering. This parameter can be even negative in the case of population inversion and then indicates the distance of $e^{+1}$ amplification.

Another kinetic parameter responsible for the scattering process, and in particular for light trapping, is the scattering length, which is given by
\begin{eqnarray}
l_{\mathrm{sc}}^{-1}(\omega)&=&n_0\sigma_{\mathrm{sc}}(\omega) ,%
\nonumber\\%
\sigma_{\mathrm{sc}}(\omega)&=&\frac{\omega\omega'^3}{c^4}\frac{1}{2F_0+1}\sum_{m'\!,m,\mathbf{e}'}\int\left|\alpha_{\mu'\mu}^{(m'm)}(\omega)e'_{\mu'}e_{\mu}\right|^2d\Omega ,%
\label{Eq4}
\end{eqnarray}
where we introduced the scattering cross section $\sigma_{\mathrm{sc}}(\omega)$ with the scattering tensor, which for the scattering on a dressed atom is specified in \cite{GerasimovRandLaser}. Here $\omega,\,\mathbf{e}$ and $\omega',\,\mathbf{e}'$ are the frequencies and polarization vectors of the input and output photons respectively and the integral is over the full scattering angle. Similarly to the extinction length, this quantity critically depends on the polarization direction of the incident photon.

Light diffusion mostly depends on the elastic contribution when $\omega=\omega'$ and both $m$ and $m'$ belong the same $F_0=3$ level. With keeping only elastic contribution we can define the characteristic length associated with losses,
\begin{equation}
l_{\mathrm{ls}}^{-1}(\omega)\equiv -l_{\mathrm{g}}^{-1}(\omega)=l_{\mathrm{ex}}^{-1}(\omega)-l_{\mathrm{sc}}^{-1}(\omega) ,%
\label{Eq5}
\end{equation}
which evaluates the averaged distance that a photon travels before being lost by an inelastic scattering event. The first equality indicates that in an amplifying medium it is possible to have this quantity negative in sign and then redefine it as the gain length $l_{\mathrm{g}}(\omega)$. In such a gain medium the radiation trapping and amplification mechanisms can overcome the instability point, entering the regime of random Raman laser generation.

\begin{figure}[tp]
{$\scalebox{1}{\includegraphics*{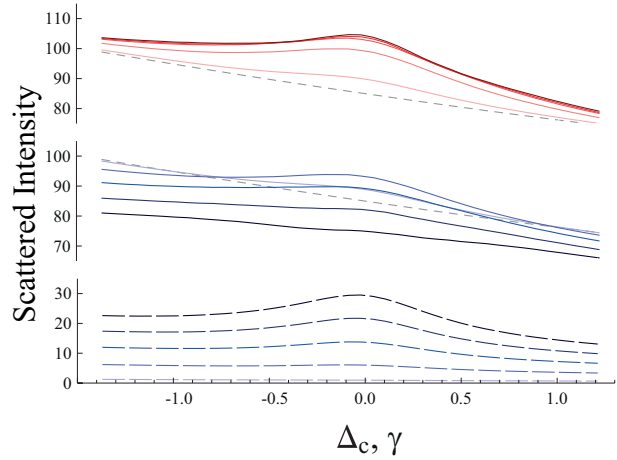}}$ }
\caption{(Color online) Intensity of the light scattered at the angle $\theta=45^\circ$ to the polarization direction of the driving fields, see Fig.~\ref{fig2}. The frequency of the optical mode is varied in the vicinity of the $F_0=2\to F=4$ forbidden transition and $\Delta_c$ is the respective detuning. The lower panel shows the contribution of the inverse anti-Stokes scattering, the middle panel is the contribution of elastic scattering, and the upper panel gives the sum for both the channels. The curve thickness in the plotted graphs is associated with the optical thickness changed from lower to higher values $b_0=1,5,10,15,20$.}
\label{fig3}%
\end{figure}

In Fig.~\ref{fig3} we show one example of the Monte-Carlo simulations from \cite{GerasimovRandLaser} for the spectral variation of the intensity of the light originally emitted by the Raman process and further scattered by the atomic ensemble under the trapping conditions. The observation channel is chosen at the angle $\theta=45^\circ$ to the polarization direction of the driving fields. In the presented data the frequency of the optical driving mode was scanned near the forbidden $F_0=2\to F=4$ transition, for particular Rabi frequencies of the control optical and microwave modes, and the optical thickness of the atomic sample on the $F_0=3\to F=4$ transition was varied from lower to higher values given by $b_0=1,5,10,15,20$. For the sake of simplicity we omit in our discussion some specific details concerning the coordination of the microwave and light induced energy shifts and the problem with spectral distribution of the emitted light \cite{GerasimovRandLaser}.

The graphs of Fig.~\ref{fig3} show that in the broad spectral domain the emitted radiation undergoes a sequence of scattering on rather complicated Autler-Townes resonance structure created by the driving fields. The susceptibility spectrum, modified by both the driving optical and microwave modes, was calculated in \cite{GerasimovRandLaser}. It is crucial that the $\chi^{(3)}$-type nonlinearity (with respect to the optical excitation) manifests itself in creation of additional quasi-energy resonances located in the atomic spectrum near the frequency of the optical mode as indicated by the dashed bar in Fig.~\ref{fig1}. The respective quasi-energy level always interacts with the modes emitted in the Raman process and traps them even if the optical mode is tuned far off-resonant from the $F =4$ reference level, i. e., in normally transparent spectral domain. But such trapping mechanism is not closed because the interaction via the quasi-energy level opens inelastic inverse anti-Stokes scattering channels as well. The anti-Stokes scattering process redistributes the Raman emission into those spectral modes, which further escape the sample without any absorption, and in the context of the random lasing effect this process should be treated as an additional loss mechanism. When the optical mode is scanned in the vicinity of the upper level $F=4$ the situation noticeably changes and the trapping effect, also initiated by elastic scattering on the closed $F_0=3\to F=4$ transition, leads to additional amplification, which is visualized as a bump-shaped enhancement of the spectra. However as follows from the presented data the significant part of the light still escapes the sample via inelastic inverse anti-Stokes scattering channel.  Although this type of scattering essentially reduces the light intensity transported via the elastic channel, when the optical mode scans the vicinity of the $F_0=2\to F=4$ transition it nevertheless stimulates Raman amplification of the trapped light, which can be seen in Fig.~\ref{fig3} in both the elastic and inelastic channels.

\section*{Raman emission and light trapping in spatially inhomogeneous conditions}

As was commented on in \cite{GerasimovRandLaser} the problem with the extra losses, related to the control mode nonlinearity, can be solved in a spatially inhomogeneous configuration. The crucial requirement for that is to separate the spatial location of the amplification and trapping areas. In the amplification area the atoms have to populate only the pumping level, i. e., $F_0=2$ in the considered case. That eliminates any inverse scattering processes. If this active area was surrounded by the atoms trapping the emitted radiation that would induce a certain soft cavity feedback and can create instability in the system.

In Fig.~\ref{fig4} we show one possible experimental architecture, which implies preparation of a spatially inhomogeneous energy structure and population distribution of the hyperfine sublevels in the atomic ensemble. This can be performed via controllable light shift of only one hyperfine energy level for the atoms located in a spatially selected volume of cylindrical symmetry inside the atomic cloud. If we assumed the level $F_0 = 2$ as light shifted and organized the population inversion for all the atoms in the selected volume onto this level (for example, with $\pi$-type microwave pulse) then these atoms could form an active medium for the photon emission. Indeed, as follows from the transition scheme, shown in Fig.~\ref{fig1}, and explaining diagrams in Fig.~\ref{fig4}, the control mode would create the photon emission on $F = 4 \to F_0 = 3$ only for the atoms inside the selected volume and the interaction would be off-resonant with the control field for the other atoms of the ensemble. Then the atoms outside of the active volume would trap the emitted light and play the role of a soft cavity redirecting the light in a quasi-one-dimensional propagation channel associated with that volume. That could lead to instability if each spontaneously emitted photon would have a diffusion path long enough for stimulation of extra photon emission while it propagates through the channel.

In Fig.~\ref{fig5} we show an example of our Monte-Carlo simulation performed for such a system under conditions demonstrating the instability behavior. These graphs track how the intensity of a single external light source located in a central point inside such a soft cavity is expanded in different orders of light scattering and depends on the amplitude of the pumping initiated by the control mode. At low pump intensity we observe an amplification of the outgoing light intensity but converging sum over a limited number of the scattering orders. At high intensity of the control mode the process becomes diverging and unstable. This effect depends on the optical density of the surrounding atoms $b_0$ and is more evident as far as the optical depth is higher. We can associate this instability with transforming the emission process towards the regime of random laser generation.

\section*{Creation of a soft cavity}

Preparation of a controlled realization of a soft cavity for this and other potential applications is a challenging experimental enterprise.  But some basic elements of the soft cavity configuration have been already demonstrated in the laboratory \cite{Itamp,Lensing}.  One important constraint is that the density of the atomic sample be relatively high, so that radiation following optical excitation within the soft cavity has a tendency to be confined within the cavity.  This suggests that one route towards experimental implementation would be to use an optical dipole trap to achieve the necessary high density.  A second constraint would be that the soft cavity should be transversely small, on the order of a few microns or less, in order to limit the number of transverse optical quasimodes within the approximately cylindrical channel geometry.

We have realized such a quasi-one-dimensional configuration by optical manipulation of a micron-scale optical dipole trap (FORT) containing ${}^{87}$Rb atoms.  The dipole trap is loaded from a magneto optical trap using procedures described elsewhere \cite{Itamp,Lensing}.  The atomic samples so produced have a peak density $\sim$ 2 $\times$ 10$^{13}$ atoms/cm$^3$ and a characteristic temperature around 100 $\mu K$. A soft cavity is generated by using a combination of optical excitation  on hyperfine components of the D$_2$ and D$_1$ lines of ${}^{87}$Rb. Two optical beams are used; the beams are combined in a single mode optical fiber, and focused tightly and cross wise to the long axis of the optical dipole trap.  One of the beams is a control beam, with a peak power on the order of 100 mW, and is tuned in the vicinity (1-100 GHz) of the D$_1$ transition around 795 nm.  We term this beam the light shift laser.  The second beam is tuned close to the nearly closed $F_0=2\to F=3$ hyperfine transition associated with the D$_2$ line at a wavelength around 780 nm.  The intensity of this probe beam is very low, well lower than the on-resonance saturation parameter for the probed transition.  The transmission of this beam through the atomic sample serves as a probe of the influence of the light shift laser.  In order to suppress detection of the light shift laser in the forward direction, several optical interference filters are used; these transmit a convenient fraction of the probe beam, while reducing the light shift laser beam intensity to undetectable levels.  We point out that it is essential to also filter the output of the lasers generating the probe and light shift laser beams so that broadband fluorescence from the diode outputs do not influence the measurements.   Additional interference filters and colored glass filters are used for this purpose.

The scheme works as follows: the probe and light shift laser beams propagate together through the atomic sample.  The light shift laser shifts energetically the atomic hyperfine ground sublevels by an amount proportional to the intensity of the laser and inversely with the detuning of the laser from resonance.   Because the light shift laser is tuned near the atomic D$_1$ line, it mainly shifts the ground hyperfine levels, and hyperfine components of the excited 5p $^{2}P_{1/2}$ level.  The excited hyperfine levels associated with the D$_2$ transition are only weakly influenced by the very far off resonance light shift laser. However, the transmission of the probe beam is strongly influence by the shift of the ground hyperfine sublevel; the effect is quite nonlinear, for the change in intensity appears as a result of the laser-induced shift of the atomic resonance; this quantity appears in the detuning dependence of the Beer's Law exponent (the optical depth), and has a strong influence on the transmitted probe intensity.

The effect is qualitatively illustrated in Fig. 6.  The upper part of the figure is generated by first taking an image of the probe alone with a charge coupled device camera.  Then a second image is made of the probe with the FORT present, and the two images subtracted.  The FORT atoms themselves are illuminated with the magneto optical trap laser beams, making essentially a flash image of the sample. The darker spot represents the probe beam transmission profile; it is several pixels in size (the pixels are 13 $\mu m$ x 13 $\mu m$.  The additional influence of the light shift laser is evident from the lower panel in Fig. 6.   In this image, generated in the same fashion as the top panel, but with the light shift laser present, the dark spot is unresolved at the single pixel level. We estimate the spot size to be several microns, consistent with model estimates of the influence of the light shift laser on the probe transmission.  Additional details including measurements of the shift and spectral profile of the $F_0=2\to F=3$ hyperfine transition are presented elsewhere \cite{Itamp}.  Further development of this approach to creating and controlling a soft cavity in a dense atomic cloud is underway.  Possible applications include study of Anderson localization of light in a cold atomic gas and studies of optical gain and the influence of atomistic disorder in quasi-one-dimension.

\section*{Concluding remarks}

In this paper we have shown an example of the conditions where a spontaneous Raman process can reach an instability point associated with the threshold of random laser generation. At present there is no microscopic quantum theory of the random laser above threshold and in the saturation regime and the performed discussion partly explain that such theory would be not so easy to develop. There are mostly phenomenological approaches presented in the literature: some extension of a diffusion model is done in \cite{WiersLagnd}, and for more information and representative references we address the reader to reviews \cite{Hui2,Hui3,HuiCao}. The coherent and correlation properties of such radiation is a most intriguing problem.

\begin{figure}[tp]
{$\scalebox{0.5}{\includegraphics*{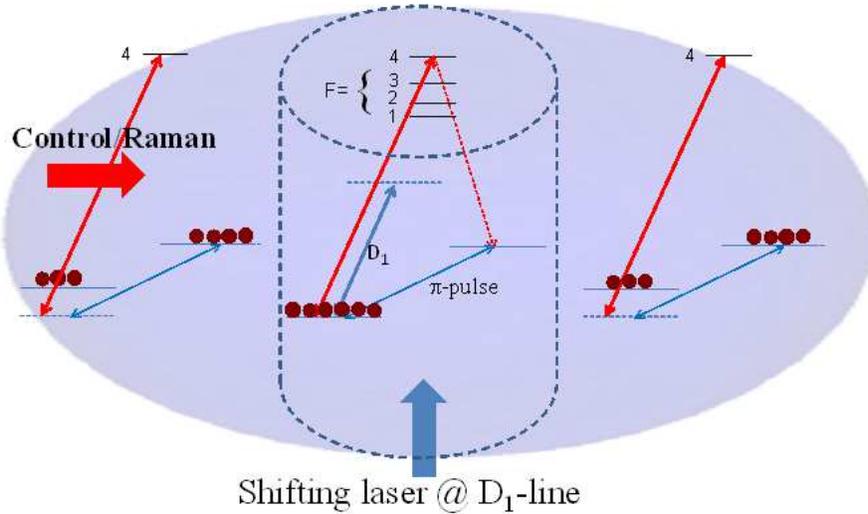}}$ }
\caption{(Color online) The excitation geometry, energy structure and transition diagram for the spatially inhomogeneous atomic system. The atoms located in the cylindrical volume crossing the middle part of the cloud have the lower energy level shifted by the light shifting laser operating near resonance with a hyperfine component of the $D_1$-transition. After applying a $\pi$-type microwave pulse these atoms populate the hyperfine transition $F_0=2$. The control mode is tuned in resonance with these atoms and initiates the initial Raman-type spontaneous emission. The emitted light is trapped by such a soft cavity and at certain conditions can enter the random lasing regime.}
\label{fig4}%
\end{figure}

\begin{figure}[tp]
{$\scalebox{0.8}{\includegraphics*{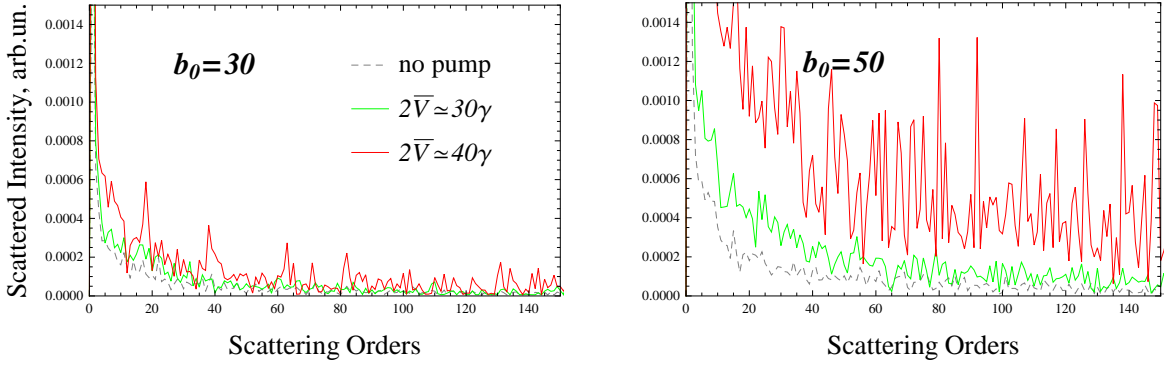}}$ }
\caption{(Color online) Intensity distribution over the scattering orders for a pointlike dipole-type light source emitting light from the center of an atomic sample. The light emerges from the trapping system along the channel shown in Fig.~\ref{fig4} and is detected at infinity at the angle of the channel direction. The gray dashed curve on both the panels refers to the intensity distribution in different scattering orders without the optical driving field, i.e.,  without amplification. Other curves show the Raman-type amplification induced by the control field (with the reduced transition matrix element $\bar{V}$ \cite{GerasimovRandLaser}) of different pump intensities with the Rabi frequencies $2\bar{V}=30 \gamma$ (green curve) and $2\bar{V}=40 \gamma$ (red curve), where $\gamma$ is the natural decay rate. In this round of the Monte-Carlo simulations we took into consideration only the incoherent scattering mechanism and ignored in our estimate additional contributions to the trapping process due to coherent scattering from the channel boundaries. Nevertheless the left- and right-hand panels, plotted for slightly different values of the sample optical depth $b_0 = 30$ and $b_0 = 50$, clearly indicate that the amplification process approaches unstable behavior for a critical optical depth. This instability can be associated with transforming the emission process towards the regime of random laser generation.}
\label{fig5}%
\end{figure}

As known from the standard theory of a conventional cavity laser \cite{Haken}, the laser radiation below threshold performs an amplified spontaneous emission of the atoms of an active medium considered as a random Langevin source. The spectral properties of the initial spontaneous emission are further modified by the amplification gain and by decay dynamics of the cavity mode. In this sense we see a certain analogy of the scheme displayed in Fig.~\ref{fig4} and the standard single mode subthreshold laser generation in a cavity. The source term performs the spontaneous emission, which would exist in any type of pumping mechanism in any laser and we emphasize that just amplification of this weak radiation source should be associated with a precursor of further laser generation. The main difference between cavity trapping and soft cavity trapping is that the latter does not select any specific spatial mode and the spectral properties of the emerging light are mostly controllable by the gain spectral profile.

Considering the process above threshold it is evident that the independent multiple scattering approach is not self-consistent to describe the feedback and wave nature of stimulated emission. The above kinetic-type approach can only track the amplification process and be applicable to identification of a steady state energy balance between the active medium and the emitted radiation. The precise theory should turn us back to the complete diagram expansion of the correlation function and to the entire Green's functions formalism substantially extended up to the saturation regime. In addition, we should also introduce a self-consistent master equation for the density matrix of atomic subsystem or more strictly for the atomic correlation functions in the complete form performed by diagram equation.

In the considered example of the Raman laser the source generation originally occurs only in a narrow spectral domain in the radiation spectrum associated with the rate of spontaneous Raman emission and in a spatial directions along a cylindrical cavity shown in Fig.~\ref{fig4}. If the amplification process, described by a Bethe-Salpeter-type equation (\ref{Eq2}), involves high scattering orders then it would become more effective for those modes, which are closer to the Raman resonance conditions. That would make the output radiation more monochromatic and therefore more coherent. We also expect that spatial inhomogeneity is an important requirement, that could lead to the spatial self-organization of the emission in the saturation regime. So we point out the difference between the standard radiation trapping phenomenon and random lasing effect as well as between conventional laser and random laser. The quantum nature of the state of light also seems a challenging problem to think about and potentially can turn us to novel sources of quantum light, i.e. an ordered single photon regime, correlated photon pairs, etc.

\begin{acknowledgments}

We thank K. Kemp and S.J. Roof for providing the images in Figure 6. This work was supported by RFBR (Grant Nos. 15-02-01060, 15-32-50411, 13-02-00944) and by the National Science Foundation (Grant Nos. NSF-PHY-0654226 and NSF-PHY-1068159). D.V.K. would like to acknowledge support from the External Fellowship Program of the Russian Quantum Center and L.V.G. from the Foundation  "Dynasty" and the Alferov's Foundation.
\end{acknowledgments}

\newpage

\clearpage

\begin{figure}[tp]
{$\scalebox{0.8}{\includegraphics*{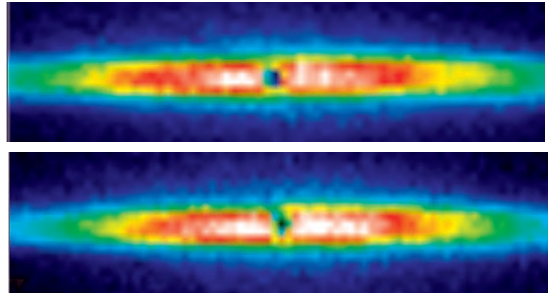}}$ }
\caption{(Color online) The upper panel illustrates absorption of a tightly focused probe beam on the $F_0=2\to F=3$ transition of ${}^{87}$Rb, as measured in forward scattering.  The lower panel shows the influence of a light shift laser on the process; the noticeably smaller dark spot, which shows the modified attenuation of the probe beam, is several microns in diameter.  }
\label{fig6}%

\end{figure}%

\end{document}